\documentclass{iosart2c}

\usepackage{amsmath,amsthm,amssymb,amscd,mathrsfs,epsfig}
\usepackage{graphicx}
\usepackage{enumerate}

\usepackage[T1]{fontenc}
\usepackage{times}%
\usepackage{amsmath}
\usepackage{dcolumn}
\usepackage{graphics}

\newtheorem{theo}{Theorem}

\newtheorem{lemm}{Lemma}
\newtheorem{exam}{Example}

\newtheorem{defi}{Definition}
\newtheorem{rema}{Remark}

\newcolumntype{d}[1]{D{.}{.}{#1}}

\firstpage{1} \lastpage{11}

\begin{document}
\begin{frontmatter}

\title{Existence of an equilibrium for pure exchange economy with fuzzy preferences}
\runningtitle{Existence of an equilibrium for pure exchange economy with fuzzy preferences}

\author[label1]{\fnms{Xia} \snm{Zhang}},
\author[label1]{\fnms{Hao} \snm{Sun}\thanks{Corresponding author. Hao sun, Department of Applied Mathematics, Northwestern Polytechnical University, Xi'an, Shaanxi 710072, P.R.China. Tel.: +86 29 88495957; Fax: +86 29
88430033; E-mail: hsun@nwpu.edu.cn.}},
\author[label1]{\fnms{Xuanzhu} \snm{Jin}},
and
\author[label2]{\fnms{Moses Olabhele} \snm{Esangbedo}}
\runningauthor{X. Zhang et al.}
\address[label1]{School of Mathematics and Statistics, Northwestern
        Polytechnical University, \\ Xi'an, Shaanxi 710072, PR China}
\address[label2]{School of Management, Northwestern Polytechnical University,\\ Xi'an, Shaanxi 710072, China}


\begin{abstract}
This paper focuses on a new model to reach the existence of equilibrium in a pure exchange economy with fuzzy preferences (PXE-FP). The proposed model integrates exchange, consumption and the agent's fuzzy preference in the consumption set. We set up a new fuzzy binary relation on the consumption set to evaluate the fuzzy preferences. Also, we prove that there exists a continuous fuzzy order-preserving function in the consumption set under certain conditions. The existence of a fuzzy competitive equilibrium for the PXE-FP is confirmed through a new result on the existence of fuzzy Nash equilibrium for fuzzy non-cooperative games. The payoffs of all strategy profiles for any agent are fuzzy numbers in fuzzy non-cooperative games. Finally, we show that the fuzzy competitive equilibrium could be characterized as a solution to an associated quasi-variational inequality, giving rise to an equilibrium solution. 
\end{abstract}


\end{frontmatter}

\section{Introduction}
The theory of competitive equilibrium was set up by Walras \cite{Walras1874}. He established a system of simultaneous equations that described an economy, and also derived the solutions to this system at equilibrium prices and quantities of commodities. However, the first rigorous result on the existence of equilibrium was reached by Wald \cite{Wald1951}. With advances in linear programming, nonlinear analysis and game theory, some discoveries about the existence of equilibrium were made by other researchers, like Mckenzie \cite{Mckenzie1959}, Hildenbrand \cite{Hildenbrand1970}, Bewley \cite{Bewley1972}, Liu \cite{Liu2017} and so on. In particular, Arrow and Debreu \cite{Arrow1954} considered the application of fixed point theory to equilibrium problems, generalizing Nash's theorem on the existence of equilibrium points for non-cooperative games \cite{Nash1950}, and then built the existence of an equilibrium in an abstract economy which is a variation on the notion of a non-cooperative game. Furthermore, an alternative approach to the study of equilibrium was given using a suitable equivalent variational inequality, such as that in Donato et al. \cite{Donato2008,Donato2016}, Anello et al. \cite{Anello2010, Anello2012}, Jor\'{e} et al. \cite{Jofre2007} and Milasi \cite{Milasi2014}.

A pure exchange economy is one without production. Each agent starts with an initial commodity bundle for trading, and has a definite order of preference on the set of all commodity bundles. Moreover, each agent's order of preference is described by a real utility function, which he acts to maximize, assuming that the prices paid and received are independent of his own choices. The concept of competitive equilibrium, as introduced by Aumann \cite{Aumann1964}, is a state of the market abiding by ``the law of supply and demand'', consisting of a price structure where the total supply of each good exactly balances the total demand and an allocation that results from trading at these prices.

It is worth noting that preference can be imagined as an individual's attitude toward a set of consumption vectors in the economy, especially as an explicit decision-making process. That is, the agent's satisfaction degree of one consumption vector relative to another (satisfaction degree of consumption vector for short) is either $0$ or $1$. According to the conclusion in Debreu \cite{Debreu1954}, there exists a real utility $a$ of any consumption vector for an agent, assuring that he has a clear-cut attitude. 
However, an agent's attitude is not necessarily clear or coherent when facing a variety of alternative consumption vectors.
For this case, Blin \cite{Blin1974} showed that the agent's satisfaction degree of any consumption vector belongs to the closed interval $[0,1]$. In such a situation, the above real utility is no longer reasonable. As a result, any agent $i$'s utility of any consumption vector can be expressed in the form of closed interval, i.e., $i$'s utility of any consumption vector which is given by a lower bound $\underline{a}$ and an upper bound $\overline{a}$. In other words, the satisfaction degree of any consumption vector for any agent is a constant in the closed interval $[0,1]$ for any utility from $\underline{a}$ to $\overline{a}$.

This research problem can be viewed from two aspects. The first aspect is vagueness in the agent's preference, which reflects the agent's indefinite satisfaction degree to many alternative consumption vectors in economics. In a case where the agent $i$'s utility of any consumption vector is given by lower and upper bounds, it is natural to suppose that $i$'s utility value falls into $[\underline{a},\overline{a}]$ following from an increasing satisfaction degree of the consumption vector. Hence, any agent's utility of any consumption vector becomes a fuzzy number $\tilde{a}$. This immediately leads us to a challenging problem of determining the agent's preference if his utility for any consumption vector is a fuzzy number. Actually, under the above assumption about an agent's satisfaction degree of any consumption vector, the satisfaction degree is not a constant value in $[0,1]$ but varies continuously in $[0,1]$. For the case where the agent's satisfaction degree of consumption vector monotonically increases with respect to his utility from lower bound to upper bound, we propose a fuzzy preference of an agent for any two consumption vectors in evaluating the degrees of relative satisfaction. Thus, we mainly study some relevant issues derived from fuzzy preferences in our paper.
The second aspect is the difficulty of determining market prices and the redistribution of goods after trading in a pure exchange economy with fuzzy preferences (PXE-FP). Primarily, we put forward the PXE-FP model, where an agent has an initial commodity bundle for trading and a fuzzy preference on the set of all commodity bundles. On the basis of this model, there are three key problems: It is difficult to evaluate the utility of different consumption vectors to agents while taking account of the fuzzy preference; it is difficult to determine whether a fuzzy competitive equilibrium exists; and it is difficult to provide a solution to compute the market equilibrium. In fact, traditional methods such as the fixed point theorem cannot provide a solution to compute the market equilibrium. Moreover, the market prices and the redistribution of goods for the PXE-FP reached by ``the law of supply and demand'', constitute the fuzzy competitive equilibrium.

Consequently, the main problem an agent confronts in a PXE-FP is choosing one or more consumption vectors from his budget set. The budget set is the set of admissible commodity vectors that an agent can afford at prices with the value of his initial endowment. Thus, a selection criterion is necessary for the agent. One approach to formalize the criterion is to suppose that the agent has a fuzzy utility index, that is, to define a fuzzy-valued function on the set of consumption vectors. It is assumed that the agent would fuzzily prefer one consumption vector to another if his fuzzy utility of one is greater than that of the other, and would be fuzzily indifferent if the fuzzy utilities of the two vectors are equal. A total order relation of fuzzy numbers is needed to compare the fuzzy utilities of different consumption vectors by which to address the agent's problem by finding all the consumption vectors that maximize the fuzzy utility on his budget set.

In this paper, we provide solutions to the three problems when considering fuzzy preferences. Firstly, as mentioned earlier, it is essential to demonstrate that the agent's fuzzy preference or indifference is represented by a fuzzy utility function that maps the consumption set onto the set of fuzzy numbers. In order to prove the conclusion, a fuzzy binary relation for any two elements in a reference set (usually a consumption set in economics) is formulated to evaluate the fuzzy preference or indifference for those. The set of fuzzy indifference classes in a reference set is defined as a quotient set. Eventually, the preceding conclusion, which enables each agent to choose a consumption vector based on the value of his fuzzy utility, is drawn from the existence of a fuzzy order-preserving function constructed by induction on the rank of the elements of the quotient set. A total order relation defined by Zhang et al. \cite{Zhang2019} using the expected values which are the centers of the expected values of interval random sets generated by these fuzzy numbers plays an important role in searching for the best consumption vector, i.e., finding out the maximal fuzzy utility.

Secondly, after developing a link between the agent's fuzzy preference or indifference and the fuzzy utility function, we aim to establish the existence of a fuzzy competitive equilibrium that provides market prices and redistribution of goods for the PXE-FP. Only based on the total order relation of fuzzy numbers and the expected mapping of the fuzzy utility function, the Kakutani's theorem \cite {Kakutani1941} can apply to prove the existence of a fuzzy Nash equilibrium for fuzzy non-cooperative games, in which the payoffs of all strategy profiles for any agent are fuzzy numbers. Consequently, we generalize the fuzzy Nash equilibrium and then prove that the fuzzy competitive equilibrium exists under some assumptions.

Thirdly, the variational inequality theory formulates an alternative approach to explicate the economic equilibrium, whose significance lies in the analysis of the properties for the equilibrium price and  allocation. However, the fuzzy preference adds to the difficulty of maximizing the fuzzy utility function. Here the expected utility function can be defined according to the expected value of the fuzzy utility for every consumption vector. Finally, by maximizing the expected utility of each agent, we can characterize the fuzzy competitive equilibrium as the solution to a related quasi-variational inequality, which results in the alternative existence of the fuzzy competitive equilibrium. As an application, an example of the PXE-FP with two goods and two agents is provided.

The motivation of this work is to establish a new fuzzy preference that is more accordant with the agent's vague attitude. Our goal is to apply the fuzzy preference to the pure exchange economy, i.e., consider the model of PXE-FP and then confirm the existence of the fuzzy competitive equilibrium of the PXE-FP. Unfortunately, neither the uniqueness nor the stability of the fuzzy competitive equilibrium is investigated in this paper.
The latter research would take into account the dynamic model of PXE-FP. It is the task of the dynamic model to show the determination of the equilibrium values of given variables under postulated conditions with various data being specified. In a real system, the discrete-time system often appears when only discrete data are available for use. Many discoveries about the discrete-time system were made by some researchers, see for example Dassios \cite{Dassios2012}, Dassios and Kalogeropoulos\cite{Dassios2013}, Oliva et al. \cite{Oliva2014}, Abraham and Kulkarni \cite{Abraham2019} and Moysis and Mishra \cite{Moysis2019}. Hence, future research will focus on the discrete-time system with fuzzy dynamic PXE-FP.

At this juncture, the main contribution of this paper is to propose a fuzzy preference and then prove that there exists a continuous fuzzy order-preserving function (utility) on the consumption set under certain conditions based on the total order relation of fuzzy numbers. The rest of this paper is presented as follows: In Section 2, we recall some basic concepts of fuzzy numbers and fuzzy mapping. Section 3 proposes the fuzzy preference relation and illustrates the link between the fuzzy preference relation and the fuzzy order-preserving function. The PXE-FP is introduced and the existence of a fuzzy competitive equilibrium is proved in Section 4. The last section concludes with a brief summary.

\section{Preliminaries}

This introduction to the theory of fuzzy sets \cite{Zadeh1965,Dubois1980,Heilpern1992} is intended to show how to define the PXE-FP.
\subsection{Fuzzy numbers}

Denote the set of all real numbers by $R$. A \emph{fuzzy set} is a mapping $\tilde{A}:R\rightarrow[0,1]$ where $\tilde{A}(x)$ assigns to each point in $R$ a \emph{grade of membership}. A \emph{fuzzy number} we treat in this paper is a fuzzy set which is upper semi-continuous, convex, normal and has bounded support. In other words, a fuzzy number is a mapping $\tilde{A}:R\rightarrow[0,1]$ with the following properties:
 \begin{enumerate}[(i)]
   \item $\tilde{A}$ is \emph{upper semi-continuous};
   \item $\tilde{A}$ is \emph{convex}, i.e.,\ $$\tilde{A}(\lambda x+(1-\lambda)y)\geq \min\{\tilde{A}(x),\tilde{A}(y)\}$$ for all $x,y\in R,\lambda \in[0,1]$;
   \item $\tilde{A}$ is \emph{normal}, i.e., $\exists \  x_{0}\in R$ for which $\tilde{A}(x_{0})=1$; and
   \item $\hbox{supp} \tilde{A}=\{x\in R\mid \tilde{A}(x)>0\}$ is a support of $\tilde{A}$ and its closure $cl(\hbox{supp}\ \tilde{A})$ is \emph{compact},
 \end{enumerate}
Let $\mathcal{F}(R)$ be the set of all fuzzy numbers in $R$.

For any $\tilde{A}\in \mathcal{F}(R)$, there exist $a,b,c,d\in R$, $\mathbb{L}:[a,b]\rightarrow[0,1]$ non-decreasing and $\mathbb{R}:[c,d]\rightarrow[0,1]$ non-increasing such that the \emph{membership function} $\tilde{A}(x)$ is given as follows:
 \begin{equation*} \label{CO}
 \tilde{A}(x)=\left\{%
  \begin{array}{ll}
  \mathbb{L}(x), &  \mbox{if}\ a\leq x< b, \\
  1, &  \mbox{if}\ b\leq x \leq c, \\
  \mathbb{R}(x), &  \mbox{if}\ c< x\leq d, \\
  0, &  \mbox{ otherwise}. \\
  \end{array}%
\right.
\end{equation*}
A fuzzy number denoted by $\lfloor a,b,c,d\rfloor$ is \emph{trapezoidal} if the functions $\mathbb{L}$ and $\mathbb{R}$ are linear.

The \emph{$\alpha$-level set} of a fuzzy number $\tilde{A}\in\mathcal{F}(R),0\leq\alpha\leq1$, denoted by $\tilde{A}[\alpha]$, is defined as
$$
\tilde{A}[\alpha]=\left\{
 \begin{array}{ll}
  \{x\in R|\tilde{A}(x)\geq\alpha \}, & \hbox{\ \mbox{if}\ $0<\alpha\leq 1,$} \\
  cl(\mbox{supp}\ \tilde{A}), & \hbox{\ \mbox{if}\ $\alpha=0.$} \\
  \end{array}
\right.
$$
It is clear that the $\alpha$-level set of a fuzzy number is a closed bounded interval $[A_{*}(\alpha),A^{*}(\alpha)]$, where $A_{*}(\alpha)$ and $A^{*}(\alpha)$ denote the left-hand and right-hand endpoint of  $\tilde{A}[\alpha]$, respectively.
Let $\tilde{A}$,\ $\tilde{B}$ be two fuzzy numbers and $\lambda$ a real number. The \emph{fuzzy addition} $\tilde{A}\tilde{+}\tilde{B}$ and \emph{scalar multiplication} $\lambda \tilde{B}$ are fuzzy numbers that have the membership functions $(\tilde{A}\tilde{+}\tilde{B})(z)$ and $\lambda\tilde{A}(z)$, defined as follows: for any $z\in R$,
$$ (\tilde{A}\tilde{+}\tilde{B})(z)=\sup_{y \in R}\{\min(\tilde{A}(y),\tilde{B}(z-y))\}, $$
$$ (\lambda\tilde{A})(z)=\left\{
 \begin{array}{ll}
  \tilde{A}(\frac{z}{\lambda}), & \hbox{\ \mbox{if}\ $\lambda\neq0,$} \\
  0, & \hbox{\ \mbox{if}\ $ \lambda=0.$} \\
  \end{array}
\right.
$$
Moreover, the $\alpha$-level sets of the fuzzy addition and the scalar multiplication have the following properties:
$$ (\tilde{A}\tilde{+}\tilde{B})[\alpha]=[ A_{*}(\alpha)+ B_{*}(\alpha), A^{*}(\alpha)+B^{*}(\alpha)],$$
$$(\lambda\tilde{A})[\alpha]=[\lambda A_{*}(\alpha),\lambda A^{*}(\alpha)],\ \mbox{if $\lambda>0$},$$
$$(\lambda\tilde{A})[\alpha]=[\lambda A^{*}(\alpha),\lambda A_{*}(\alpha)],\ \mbox {if $\lambda<0$},$$
$$(\tilde{A}\tilde{+} \lambda)[\alpha] =(\tilde{A}\tilde{+}\tilde{\lambda})[\alpha]=[A_*(\alpha)+\lambda,A^*(\alpha)+\lambda].$$

The \emph{expected value} $E(\tilde{A})$ of a fuzzy number $\tilde{A}$ is defined as follows:
$$ E(\tilde{A})=\tfrac{1}{2}\int_0^1(A_{*}(\alpha)+A^{*}(\alpha))d\alpha.$$

For any $\tilde{A},\tilde{B}\in\mathcal{F}(R)$, the expected values of fuzzy numbers satisfy the following properties:
$$ E(\tilde{A}\tilde{+}\tilde{B})=E(\tilde{A})+E(\tilde{B}), E(\tilde{A}\tilde{-}\tilde{B})=E(\tilde{A})-E(\tilde{B}).$$

Using the expected values, a \emph{total order relation} of fuzzy numbers was introduced by \cite{Zhang2019}, that is, for any $\tilde{A},\tilde{B}\in\mathcal{F}(R)$, we say $\tilde{A}$ is weakly superior to $\tilde{B}$, denoted by $\tilde{A}\succcurlyeq\tilde{B}$, if and only if $E(\tilde{A})\geq E(\tilde{B})$; $\tilde{A}$ and $\tilde{B}$ are an indifference relationship, denoted by $\tilde{A}\thickapprox\tilde{B}$, if and only if $E(\tilde{A})=E(\tilde{B})$; $\tilde{A}$ is superior to $\tilde{B}$, denoted by $\tilde{A}\succ\tilde{B}$, if and only if $ E(\tilde{A})>E(\tilde{B})$.

From the definition of the total order relation of fuzzy numbers, it is easy to show that for any subset $\tilde{X}$ of the set of all fuzzy numbers, i.e., $\tilde{X}\subseteq\mathcal{F}(R)$, the \emph{maximum value} and \emph{supremum} of the set $\tilde{X}$ are defined as:
$$\tilde{C}\approx\max_{\tilde{A}\in \tilde{X}}\tilde{A}\Leftrightarrow E(\tilde{C})=\max_{\tilde{A}\in\tilde{X}}E(\tilde{A}),$$
$$\tilde{C}\approx\sup_{\tilde{A}\in \tilde{X}}\tilde{A}\Leftrightarrow E(\tilde{C})=\sup_{\tilde{A}\in \tilde{X}}E(\tilde{A}),$$
The \emph{minimum value} and \emph{infimum} of the set $\tilde{X}$ are defined in the same way.

\subsection{Fuzzy mapping}

In what follows, for any $\mathbf{x}\in R^l$ and $\delta>0$, let $B_{\delta}(\mathbf{x})=\{\mathbf{y}\in R^l\mid \|\mathbf{y}-\mathbf{x}\|<\delta\}$. Then, in succession, we give some related concepts of fuzzy mapping.
\begin{defi}
\emph{
Let $X$ be the non-empty subset of $R^l$. A \emph{fuzzy mapping} $\tilde{f}:X\rightarrow\mathcal{F}(R)$ is said to be:
\begin{enumerate}[(i)]
\item \emph{upper semicontinuous} at $\mathbf{x}_0\in X$ if for any $\tilde{\varepsilon}=\varepsilon>0$ there exists a $\delta=\delta(\mathbf{x}_0,\varepsilon)>0$ such that
\begin{align}\label{AB}
& \tilde{f}(\mathbf{x})\preccurlyeq\tilde{f}(\mathbf{x}_0)\tilde{+}\tilde{\varepsilon}
\end{align}
for all $\mathbf{x}\in X\cap B_{\delta}(\mathbf{x}_0)$, and
$\tilde{f}:X\rightarrow\mathcal{F}(R)$ is upper semicontinuous if it is upper semicontinuous at any point of $X$;
\item \emph{lower semicontinuous} at $\mathbf{x}_0\in X$ if for each $\tilde{\varepsilon}=\varepsilon>0$ there exists a $\delta=\delta(\mathbf{x}_0,\varepsilon)>0$ such that
\begin{align}\label{AC}
& \tilde{f}(\mathbf{x}_0)\preccurlyeq\tilde{f}(\mathbf{x})\tilde{+}\tilde{\varepsilon}
\end{align}
for all $\mathbf{x}\in X\cap B_{\delta}(\mathbf{x}_0)$, and
$\tilde{f}:X\rightarrow\mathcal{F}(R)$ is lower semicontinuous if it is lower semicontinuous at each point of $X$; and
\item \emph{continuous} at $\mathbf{x}_0\in X$ if it is upper semicontinuous and lower semicontinuous at $\mathbf{x}_0\in X$.
\end{enumerate}
}
\end{defi}

Let $\tilde{f}:X\rightarrow\mathcal{F}(R)$ be a fuzzy mapping parameterized by
$$\tilde{f}(\mathbf{x})=\{(f(\mathbf{x})_*(\alpha),f(\mathbf{x})^*(\alpha),\alpha):\alpha \in[0,1]\},$$
for each $\mathbf{x}\in X$.

The expected mapping $f_E(\mathbf{x})$ for any $\mathbf{x}\in X$ defined as
$$f_E(\mathbf{x})=\frac{1}{2}\int_0^1[f(\mathbf{x})_{*}(\alpha)+f(\mathbf{x})^{*}(\alpha)]d\alpha,$$
is a real-valued function. Consequently, (\ref{AB}) and (\ref{AC}) can be written as
\begin{align*}
& f_E(\mathbf{x})\leq f_E(\mathbf{x}_0)+\varepsilon
\end{align*}
and
\begin{align*}
& f_E(\mathbf{x}_0)\leq f_E(\mathbf{x})+\varepsilon.
\end{align*}
Following from the expected mapping, we can get the following result which describes a continuous fuzzy function as a continuous real function.
\begin{theo}
Let $\tilde{f}:X\rightarrow\mathcal{F}(R)$ be a fuzzy mapping parameterized by
$$\tilde{f}(\mathbf{x})=\{(f(\mathbf{x})_*(\alpha),f(\mathbf{x})^*(\alpha),\alpha):\alpha \in[0,1]\},$$
for each $\mathbf{x}\in X$.
The fuzzy mapping $\tilde{f}$ is continuous at $\mathbf{x}_0\in X$ if and only if its expected mapping is continuous at $\mathbf{x}_0$.
\end{theo}

\begin{defi}
\emph{
Let $X$ be a non-empty convex subset of $R^l$. $\tilde{f}:X\rightarrow\mathcal{F}(R)$ is said to be \emph{fuzzy quasi-concave} if for any $\mathbf{x},\mathbf{y}\in X$,
\begin{align}\label{AH}
& f_E(\lambda \mathbf{x} +(1-\lambda)\mathbf{y})\geq \min\{f_E(\mathbf{x}),f_E(\mathbf{y})\},
\end{align}
for each $\lambda\in (0,1)$,
where $f_E(\mathbf{x})$ is the expected mapping of $\tilde{f}(\mathbf{x})$. Moreover, $\tilde{f}$ is said to be \emph{strictly fuzzy quasi-concave} if inequality (\ref{AH}) strictly holds for $ f_E(\mathbf{x})\neq f_E(\mathbf{y})$.
}
\end{defi}

\section{Fuzzy preference relation and fuzzy order-preserving function}

Generally, preference is strongly linked to an individual's explicit attitude toward a collection of objects that can influence his decision making. For instance, preference implies the agent's satisfaction degree of any consumption vector, which is either $0$ or $1$ in the economy. Moreover, there is a real utility of any consumption vector for an agent verified by Debreu \cite{Debreu1954}. In fact, an agent's attitude is ambiguous when facing a variety of alternative consumption vectors. In other words, his choice from the consumption set is not normally in line with his preference $\succ^i$, i.e., the binary relation ($0$ or $1$) with respect to any two consumption vectors is not an adequate explanation for his attitude. A different binary relation was introduced by Nakamura \cite{Nakamura1986}.

Let $X$ be a reference set. A binary relation $\mathfrak{R}$ of $X$, defined by Nakamura \cite{Nakamura1986}, is characterized by a membership function:
$$\mu_{\mathfrak{R}}:X\times X\rightarrow [0,1].$$

Under this circumstance, an agent's utility of any consumption vector can be given by a lower bound and an upper bound, which means the satisfaction degree of any consumption vector for an agent is a constant in $[0,1]$ for any utility from lower bound to upper bound. However, naturally assume that the agent's satisfaction degree of any consumption vector monotonically increases with regard to his utility from lower bound to upper bound. Consequently, the agent's satisfaction degree of any consumption vector is not a constant value in $[0,1]$ but varies continuously in $[0,1]$. Therefore, it is necessary to define the following fuzzy binary relation $\mathcal{G}$ of a reference set $X$ (usually in the finite vector space of commodity bundles in economics).

\begin{defi}\label{01}
\emph{
Let $X$ be a reference set. A \emph{fuzzy binary relation} $\mathcal{G}$ of $X$ is characterized by a membership function
$$\mu_{\mathcal{G}}: X\times X\rightarrow \mathcal{F}(R).$$
}
\end{defi}

Based on the fuzzy binary relation $\mathcal{G}$, we define a fuzzy preference relation $\succsim_{\mathcal{G}}$ on a reference set $X$.

\begin{defi}\label{C19}
\emph{
For any $ x,y\in X$, if $\mu_{\mathcal{G}}(x,y)\succcurlyeq \mu_{\mathcal{G}}(y,x)$, we say $x$ is \emph{fuzzily weakly preferred} to $y$, denoted by $x\succsim_{\mathcal{G}} y$; if $\mu_{\mathcal{G}}(x,y)\approx \mu_{\mathcal{G}}(y,x)$, $x$ is \emph{fuzzily indifferent} to $y$, denoted by $x\sim_{\mathcal{G}} y$;
$x$ is \emph{fuzzily preferred} to $y$, denoted by $x\succ_{\mathcal{G}} y$, if $\mu_{\mathcal{G}}(x,y)\succ \mu_{\mathcal{G}}(y,x)$.
}
\end{defi}

Observe that the fuzzy preference relation $\succsim_{\mathcal{G}}$ is deemed to be ``consistent'' if $\mu_{\mathcal{G}}(x,y)\succcurlyeq \mu_{\mathcal{G}}(y,x)$ and $\mu_{\mathcal{G}}(y,z)\succcurlyeq \mu_{\mathcal{G}}(z,y)$ imply that $\mu_{\mathcal{G}}(x,z)\succcurlyeq \mu_{\mathcal{G}}(z,x)$.

We assume the fuzzy preference relation $\succsim_{\mathcal{G}}$ is ``consistent'' in this paper. Owing to the total order relation of fuzzy numbers defined by Zhang et al. \cite{Zhang2019}, the fuzzy preference relation $\succsim_{\mathcal{G}}$ of a reference set $X$ satisfies the following properties:
\begin{enumerate}[(1)]
\item For any $x\in X$, $x\succsim_{\mathcal{G}} x$;
\item For any $x,y,z\in X$, $x\succsim_{\mathcal{G}} y$, and $y\succsim_{\mathcal{G}} z$, it yields that $x\succsim_{\mathcal{G}} z$;
\item For any $x,y\in X$, $x\succsim_{\mathcal{G}} y$ and/or $y\succsim_{\mathcal{G}} x$;
\item For any $x,y\in X$, if $x\succsim_{\mathcal{G}} y$ and $y\succsim_{\mathcal{G}} x$, then $x\sim_{\mathcal{G}} y$.
\end{enumerate}
Meanwhile, we say that the fuzzy preference relation $\succsim_{\mathcal{G}}$ is a completely ordered relation and $(X,\succsim_{\mathcal{G}})$ is a completely ordered space.

Note that for any $x,y,z$ belonging to a reference set $X$, the fuzzy interval $[x,y]$ (or $(x,y)$) denotes $\{z \mid x\precsim_{\mathcal{G}}z\precsim_{\mathcal{G}}y\} $ (or $\{z\mid x\prec_{\mathcal{G}}z\prec_{\mathcal{G}} y\}$); for any $\tilde{a}, \tilde{b}, \tilde{c}\in \mathcal{F}(R)$, the fuzzy number interval $[\tilde{a},\tilde{b}]$ (or $(\tilde{a},\tilde{b})$) denotes $\{\tilde{c}\mid \tilde{a}\preccurlyeq \tilde{c}\preccurlyeq \tilde{b} \}$ (or $\{\tilde{c}\mid \tilde{a}\prec \tilde{c}\prec\tilde{b} \}$).

A \emph{completely ordered topology} is the topology generated by the fuzzy intervals.
A \emph{natural topology} on a reference set $X$ is a completely ordered topology for which the sets $\{x\in X\mid x \precsim_{\mathcal{G}} x'\}$ and $\{x\in X\mid x'\precsim_{\mathcal{G}} x\}$ are closed for all $x'\in X$, where the closed set $\{x\in X\mid x \precsim_{\mathcal{G}}x'\}$ implies that for any sequence $\{x^{(n)}\}$ of points in $X$ with a limit $x^0\in X$, if for all $n$, $x^{(n)}\precsim_{\mathcal{G}}x'$, then $x^0\precsim_{\mathcal{G}} x'$.

A fuzzy function $\tilde{f}(x): X\rightarrow \mathcal{F}(R)$ defined on a reference set $X $ is said to be \emph{order-preserving} if $ x\precsim_{\mathcal{G}}y$ is equivalent to $\tilde{f}(x)\preccurlyeq \tilde{f}(y)$. The domain of values of the function $\tilde{f}$ is denoted by $\tilde{f}(X)$.

The \emph{quotient set} $X/_{\sim}=\{q_x\mid \forall x \in X\}= Q$ is the set of all \emph{fuzzy indifference classes} in a reference set $X$ denoted by $Q$,
where $q_x=\{y\in X\mid y\sim_{\mathcal{G}}x\}$ is the collection of all elements fuzzily indifferent to $x$ in $X$. For any $q\in Q$, $q$ is a fuzzy indifference class in $X$.


\begin{lemm}\label{AI}
Given the fuzzy preference relation $\succsim_{\mathcal{G}}$ and a reference set $X$, let the quotient set $Q$ of $X$ be countable. There exists a continuous fuzzy order-preserving function in any natural topology on $X$.
\end{lemm}

\noindent\textbf{Proof.} Due to the total order relation of fuzzy numbers and the expected function of a fuzzy mapping, it is possible to construct a fuzzy order-preserving function $\tilde{\phi}$ mapping $Q$ into some finite fuzzy number interval by induction on the rank of the element of $Q$. It is assumed that there exist two fuzzy numbers $\tilde{a}$ and $\tilde{c}$ such that $\tilde{a}\approx\inf\limits_{q\in Q} \tilde{\phi}(q), \tilde{c}\approx\sup\limits _{q\in Q}\tilde{\phi}(q)$.
 If $\tilde{b}'$ satisfies $\tilde{a}\prec\tilde{b}'\prec \tilde{c}$ and $\tilde{b}'\notin \tilde{\phi}(Q)$,
the following four cases may occur:
\begin{enumerate}[(i)]
\item the set $\{\tilde{b}\in \tilde{\phi}(Q)\mid \tilde{b}\prec\tilde{b}'\}$ may have a largest element;
\item the set $\{\tilde{b}\in \tilde{\phi}(Q)\mid \tilde{b}\prec\tilde{b}'\}$ may not have a largest element;
\item the set $\{\tilde{b}\in \tilde{\phi}(Q)\mid \tilde{b}'\prec\tilde{b}\}$  may have a smallest element;
\item the set $\{\tilde{b}\in \tilde{\phi}(Q)\mid \tilde{b}'\prec\tilde{b}\}$  may not have a smallest element.
\end{enumerate}
We intend to remove the gaps of types (i-iv), (ii-iii) and (ii-iv). Let us present a fuzzy non-decreasing step function $\tilde{\psi}(\tilde{b})$ for which the height of each step is equal to the length of the corresponding gap. Then the new fuzzy function $\tilde{f}^0(q)=\tilde{\phi}(q)\tilde{-}\tilde{\psi}[\tilde{\phi}(q)]$ which maps $Q$ into some finite fuzzy number interval is still order-preserving and $\tilde{f}^0(Q)$ has no gaps of the unwanted types. We define $\tilde{f}(x)=\tilde{f}^0(q_{x})$. Aiming to show that $\tilde{f}$ is continuous in any natural topology on $X$, we have to consider a fuzzy number $\tilde{b}'$ so that $ \tilde{a}\prec\tilde{b}'\prec\tilde{c}$ and the set $X_{\tilde{b}'}=\{x\in X\mid \tilde{f}(x)\preccurlyeq\tilde{b}'\}$.
\begin{enumerate}[(a)]
 \item If $\tilde{b}' \in \tilde{f}(X)$, let $x'\in X$ be such that $\tilde{b}'\approx\tilde{f}(x')$, which means $X_{\tilde{b}'}=\{x\in X\mid \tilde{f}(x) \preccurlyeq \tilde{f}(x')\}$. Because $\tilde{f}$ is order-preserving, it implies $X_{\tilde{b}'}=\{x\in X\mid x \precsim_{\mathcal{G}}x'\}$ and thus is a closed set.
\item Supposing that $\tilde{b}' \notin \tilde{f}(X)$ and the set $W_{\tilde{b}'}=\{\tilde{b}\in \tilde{f}(X)\mid\tilde{b}\prec\tilde{b}'\}$ has a largest element $\tilde{b}''$, this means $X_{\tilde{b}''}=X_{\tilde{b}'}$ which is closed by (a).
\item With the hypotheses that $\tilde{b}' \notin \tilde{f}(X)$ and the set $W_{\tilde{b}'}$ has no largest element, it involves the set $W^{\tilde{b}'}=\{\tilde{b}\in \tilde{f}(X)\mid\tilde{b}'\prec\tilde{b}\}$ without a smallest element since $\tilde{f}(X)$ has no gap of type (ii-iii). Thus $X_{\tilde{b}'}= \bigcap\limits_{\tilde{b}\in W^{\tilde{b}'}} X_{\tilde{b}}$ and $X_{\tilde{b}'}$ is closed as an intersection of closed sets.
\end{enumerate}
In the same way, the verdict holds for the set $X^{\tilde{b}'}=\{x\in X\mid \tilde{b}'\preccurlyeq \tilde{f}(x)\}$ of any fuzzy number $\tilde{b}'$. It follows that the inverse image by $\tilde{f}$ of any closed set is a closed set on $X$. \qed

\begin{lemm}\label{AJ}
Given the fuzzy preference relation $\succsim_{\mathcal{G}}$ and a reference set $X$, let $\mathcal{R}$ be a countable subset of $X$. For every pair $x,y\in X$ meeting $x \precsim_{\mathcal{G}} y$, if there is an element $r$ of $\mathcal{R}$ such that $x\precsim_{\mathcal{G}} r\precsim_{\mathcal{G}}y$, then there exists a continuous fuzzy order-preserving function in any natural topology on $X$.
\end{lemm}
\noindent\textbf{Proof.}  Let us consider the two quotient sets $X/_{\sim}=Q$ and $\mathcal{R}/_{\sim}=D$. $D$ is countable. If $Q$ has a smallest and/or a largest element, without any loss of generality, we can suppose that both possible elements belong to $D$.

Here, we define another equivalence relation among the elements of $Q$ as: $q_1Fq_2$ if and only if there exists a finite number of elements of $Q$ between $q_1$ and $q_2$. Also, equivalence classes for $F$ are denoted by $[q_1]_F,[q_2]_F,\cdots,[q_r]_F$,
$\cdots$.
Every equivalence class is countable. Moreover an equivalence class $[q]_F$ with more than one element of $Q$ contains an element of $D$, which implies that the equivalence classes $[q]_F$ form a countable set. We denote the union over these classes $[q]_F$ by $D'$ which is countable and then define $T=D\cup D'$.

As in the proof of Lemma \ref{AI}, we construct the function $\tilde{f}^0$ on $T$ and extend it from $T$ to $Q$ as follows. Let $q\in Q$ and $ q\notin T$. It is found that the set $T_q=\{t\in T\mid t\prec_{\mathcal{G}} q\}$ has no largest element. In fact, for any $t'\in T_q$, $q\notin T$ so that $q\notin D'$. Besides, there is an infinity of elements of $Q$ between $t'$ and $q$, and then there exists an infinity of elements of $T$ between $t'$ and $q$. Similarly the set $T^q=\{t\in T\mid q\prec_{\mathcal{G}} t\}$ has no smallest element. That means the value $\sup\limits_{t\in T_q}\tilde{f}^0(t)$ equals to $\inf \limits_{t\in T^q}\tilde{f}^0(t)$, since $\tilde{f}^0(T)$ has no gap of type (ii-iv). This equal value defines $\tilde{f}^0(q)$, which is clearly order-preserving, and therefore the fuzzy function $\tilde{f}(x)=\tilde{f}^0(q_{x})$ is order-preserving. Furthermore, analogous to the proof of Lemma \ref{AI}, we can conclude that $\tilde{f}(x)$ is continuous in any natural topology on $X$, since $\tilde{f}(X)=\tilde{f}^0(Q)$ has no gap of type (i-iv) or (ii-iii).\qed

A completely ordered topological space $(X,\succsim_{\mathcal{G}})$ is perfectly separable if there exists a countable class for any open set in $(X,\succsim_{\mathcal{G}})$ such that the open set is the union of the sets of the class.
\begin{theo}\label{C4}
Given the fuzzy preference relation $\succsim_{\mathcal{G}}$ and a reference set $X$, let $(X,\succsim_{\mathcal{G}})$ be a perfectly separable space. If for every $x'\in X$, the sets $\{x\in X\mid x\precsim_{\mathcal{G}}x'\}$ and $\{x\in X\mid x'\precsim_{\mathcal{G}} x\}$ are closed, there exists a continuous fuzzy order-preserving function on $X$.
\end{theo}
\noindent\textbf{Proof.} We can choose an element in any non-empty set $S$ of $X$. In this way, these elements form a countable set $\mathcal{R}''$.
Let us consider the pair $q_1,q_2\in Q= X/_\sim$ satisfying the conditions that $q_1\prec_{\mathcal{G}}q_2$ and there does not exist a fuzzy indifference class $q_3\in Q$ such that $q_1\prec_{\mathcal{G}}q_3 \prec_{\mathcal{G}}q_2$. Assert that the set of those pairs is countable. To prove this, we take two elements $x',y'$ in the indifference classes $q_1$ and $q_2$ respectively. Moreover, the set $\{x\in X\mid x\prec_{\mathcal{G}}y'\}$ is open and hence there exists a set $S_{q_2}$ in the class of $S$ such that $x'\in S_{q_2}\subseteq\{x\in X\mid x\prec_{\mathcal{G}}y'\}$. Provided that $q_1',q_2'$ is another pair possessing the same properties, $S_{q_2'}$ is different from $S_{q_2}$. If $q_1\prec_{\mathcal{G}}q_2\precsim_{\mathcal{G}} q_1'\prec_{\mathcal{G}}q_2'$, then $x''\in S_{q_2'}$ and $x''\notin S_{q_2}$. Else if $q_1'\prec_{\mathcal{G}}q_2'\precsim_{\mathcal{G}}q_1 \prec_{\mathcal{G}}q_2$, then $x'\in S_{q_2}$ and $x'\notin S_{q_2'}$. It is obvious that the pair $q_1,q_2$ is in one-to-one correspondence with a subclass of the countable class of $S$. Then choose an element $x'$ in each class $q_1$ and an element $y'$ in each class $q_2$. All those $x'$ and $y'$ form a countable set named $\mathcal{R}'$.

Let us examine the countable set $\mathcal{R}=\mathcal{R}'\cup \mathcal{R}''$. It satisfies all the properties required by Lemma \ref{AJ}. Let $x,y$ be a pair of elements of $X$ such that $x \prec_{\mathcal{G}} y$. If the set $(x,y)$ is non-empty, it contains a non-empty set $S$ and therefore an element of $\mathcal{R}''$. Otherwise, $x\sim_{\mathcal{G}} x'\in \mathcal{R}'$ and $y\sim_{\mathcal{G}}y'\in \mathcal{R}'$. In any case, $[x,y]$ contains an element of $\mathcal{R}$.\qed

Finally, for a given fuzzy preference relation $\succsim_{\mathcal{G}}$, a \emph{fuzzy utility function} is an order-preserving function $\tilde{u}(x)$ that maps reference set $X$ into the set of all fuzzy numbers. Observe that, under the condition that for every $x'\in X$ the sets $\{x\in X\mid x\precsim_{\mathcal{G}}x'\}$ and $\{x\in X\mid x'\precsim_{\mathcal{G}} x\}$ are closed, the fuzzy utility function $\tilde{u}(x)$ on $X$ is continuous.

\section{The existence of a fuzzy competitive equilibrium}

In our paper, for any vectors $\mathbf{x}_i,\mathbf{y}_i\in R^l$, $\mathbf{x}_i>\mathbf{y}_i$ means $x_{ih}>y_{ih}$ for all $h$; $\mathbf{x}_i \geqq \mathbf{y}_i$ means $x_{ih} \geq y_{ih}$ for all $h$; and $\mathbf{x}_i\geqslant \mathbf{y}_i$ means $\mathbf{x}_i\geqq \mathbf{y}_i$ but not $\mathbf{x}_i=\mathbf{y}_i$. The scalar product $\sum\limits_{h=1}^l x_{ih}y_{ih}$ of two members $\mathbf{x}_i$ and $\mathbf{y}_i$ of $R^l$ is denoted by $\langle \mathbf{x}_i,\mathbf{y}_i\rangle$.

\subsection{PXE-FP model}

In the classical pure exchange economy, we take into account a marketplace consisting of $l$ different goods indexed by $h=1,\cdots,l$ and $m$ agents denoted by $i=1,\cdots,m$. Every agent $i$ has an initial endowment vector: $\mathbf{w}_i=(w_{i1},\cdots,w_{il})\in R^l_+$. The consumption vector relative to the agent $i$ is $\mathbf{x}_i=(x_{i1},\cdots,x_{il})\in X_i\subseteq R^l_+$, where $x_{ih}$ is consumption relative to the commodity $h$, $X_i$ is interpreted as the consumption set of agent $i$, and $\mathbf{x}=(\mathbf{x}_1,\cdots,\mathbf{x}_m)\in \prod\limits_{i=1}^m X_i$ represents the consumption of the market. We denote the price vector by $\mathbf{p}=(p_1,\cdots,p_l)\in R^l_+$, where $p_h (h=1,\cdots,l)$ is the price of commodity $h$. As is standard in economic theory, the choice by the agent from a given set of alternative consumption vectors is assumed to be made in accordance with his preference $\succ^i$. The reason is that there exists a utility function: $u_i:R^l_+\rightarrow R$ such that $u_i(\mathbf{x}_i)\geq u_i(\mathbf{x}'_i)$ if and only if $\mathbf{x}_i$ is preferred or indifferent to $\mathbf{x}'_i$ for agent $i$.

Nevertheless, the agent's attitude is ambiguous when facing all sorts of alternative consumption vectors, which is analyzed in Section 3. Accordingly, we define a fuzzy preference relation $\succsim_{\mathcal{G}}$ on a reference set as Definition \ref{C19}, in order to better propose a new model which we study in this paper.
In this section, an agent $i$'s fuzzy preference relation on a consumption set $X_i$ is denoted by $\succsim^i_{\mathcal{G}}$.

\begin{defi}
\emph{
A \emph{PXE-FP} is defined as
$$\tilde{\mathcal{E}}=(R_+^l,X_i,\succsim^i_{\mathcal{G}},\mathbf{w}_i)$$
consisting of $m$ agents indexed by $i=1,\cdots,m$, each of which has a fuzzy preference $\succsim^i_{\mathcal{G}}$ as well as an initial endowment vector $\mathbf{w}_i=(w_{i1},\cdots,w_{il})\in R^l_+$, and trades $l$ different goods denoted by $h=1,\cdots,l$, where $X_i\subseteq R_+^l$ is the consumption set of agent $i$ and its element $\mathbf{x}_i=(x_{i1},\cdots,x_{il})$ is a consumption vector of the agent.
}
\end{defi}
\begin{rema}
\emph{
The pure exchange economy is a special case of the PXE-FP when the agent's satisfaction degree for any consumption vector is only $0$ or $1$.}
\end{rema}


An agent's motivation in the choice of a consumption vector is to maximize his fuzzy utility among all consumption vectors that belong to his budget set, admissible consumption vectors of which are affordable for the agent at price vector $\mathbf{p}=(p_1,\cdots,p_l)$ with the value of his initial endowment vector $\mathbf{w}_i$. In turn, the agent's income can be regarded as the receipts from sales of the initial endowments.

$\mathbf{Condition}$ (1) $\bar{\mathbf{x}}_i$ is the optimum solution of $\max\limits_{\mathbf{x}_i\in B_i(\bar{\mathbf{p}})}\tilde{u}_i(\mathbf{x}_i)$, where $B_i(\bar{\mathbf{p}})=\{\mathbf{x}_i\mid \mathbf{x}_i\in X_i,\langle\bar{\mathbf{p}}$,
$\mathbf{x}_i\rangle\leq\langle\bar{\mathbf{p}},\mathbf{w}_i\rangle\}$, for all $i=1,\cdots,m$.\label{B3}

It is required that the prices of different goods be non-negative and not all zero. Without any loss of generality, we can normalize the vector $\bar{\mathbf{p}}$ by restricting the sum of its coordinates to be $1$.

$\mathbf{Condition}$ (2) $\bar{\mathbf{p}}\in P=\{\mathbf{p}\mid \mathbf{p}\in R^l,\mathbf{p}\geqq \mathbf{0},\sum\limits_{h=1}^l p_h=1\}$.\label{B4}

The market for any goods is usually considered to be in equilibrium if the supply of the good equals the demand for it. However, the price of some good may be zero, which means supply will exceed demand. The aggregate excess demand is $\mathbf{z}=(z_1,\cdots,z_l)\in R^l$, where $z_h=\sum\limits_{i=1}^m(x_{ih}-w_{ih})$ and $x_{ih}-w_{ih}$ is the individual excess demand of agent $i$ relative to good $h=1,\cdots,l$.

$\mathbf{Condition}$ (3) $\bar{\mathbf{z}}\leqq\mathbf{0},\langle\bar{\mathbf{p}},\bar{\mathbf{z}}\rangle=0$. \label{B5}

\begin{defi}\label{Ay}
\emph{
For PXE-FP $\tilde{\mathcal{E}}$, a pair $(\bar{\mathbf{p}}, \bar{\mathbf{x}})$ is said to be a \emph{fuzzy competitive equilibrium} of $\tilde{\mathcal{E}}$ if it satisfies Conditions (1)-(3).
}
\end{defi}

A fuzzy competitive equilibrium is a state of the market arriving at by ``the law of supply and demand", which consists of a competitive equilibrium price $\bar{\mathbf{p}}$ and a competitive equilibrium allocation $\bar{\mathbf{x}}$ such that the fuzzy utility of $\bar{\mathbf{x}}_i$ for any agent $i$ in his budget set is maximal.

In what follows, we obtain the existence result of a fuzzy competitive equilibrium from two perspectives, with the total order relation of fuzzy numbers and the expected function of a fuzzy mapping as key points. The first method mainly generalizes the existence of the fuzzy Nash equilibrium confirmed by fixed point theorem in \cite{Nash1950}. Nevertheless, the uniqueness of the fuzzy competitive equilibrium cannot be illustrated by the first method. So then, we take into account the second approach which applies an associated quasi-variational inequality. If the solution of the corresponding quasi-variational inequality is unique under some specific conditions, then there exists only one fuzzy competitive equilibrium of the PXE-FP. Furthermore, the fuzzy competitive equilibrium can be characterized by the solution of the quasi-variational inequality.

We will obtain the existence of the fuzzy Nash equilibrium for a fuzzy non-cooperative game before substantiating the existence of a fuzzy competitive equilibrium for a PXE-FP.

\subsection{Fuzzy non-cooperative games}

A \emph{classical non-cooperative game} consists of $n$ players indexed by $N=\{1,$
$\cdots,n\}$. Any player $i$ has his own set of possible strategies $S_i$ and players except $i$ have strategy profiles $S_{-i}=S_1\times\cdots\times S_{i-1}\times S_{i+1} \times\cdots \times S_n$. To play the game, under the condition that other players choose a strategy profile $s_{-i}\in S_{-i} $, player $i$ selects a strategy $s_i\in S_i$. $s=(s_i,s_{-i})\in S$ denotes the vector of strategies selected by the players and $S=\prod\limits_{i=1}^n S_i$ is the set of strategy profiles of the game. The vector of strategies $s\in S$ chosen by the players determines the payoff for each player. 
Generally, $u_i:S\rightarrow R$ is used to denote the payoff function of the $i$-th player and $G=( N,S_i,u_i)$ denotes a classical noncooperative game.

In consideration of the impression of information in decision-making problems, we define a fuzzy non-cooperative game by integrating into a fuzzy payoff.

\begin{defi}
\emph{
A \emph{fuzzy non-cooperative game} is defined as $G_{\mathcal{F}}=(N,S_i,\tilde{u}_i)$ consisting of $n$ players indexed by $i \in N=\{ 1, \cdots,n\}$, each of whom has a strategy set $S_i$ as well as a fuzzy payoff function $\tilde{u}_i:S\rightarrow \mathcal{F}(R)$, where $S=\prod\limits_{i=1}^n S_i$ is the set of strategy profiles.
}
\end{defi}

\begin{defi}\label{C6}
\emph{
For any fuzzy non-cooperative game $G_{\mathcal{F}}=(N,S_i,\tilde{u}_i)$, $(s^\star_i,s^\star_{-i})$ is said to be a \emph{fuzzy Nash equilibrium} if and only if
$$\tilde{u}_i(s^\star_i,s^\star_{-i})\succcurlyeq \tilde{u}_i(s_i,s^\star_{-i}),\ \hbox{for any} \ i\in N, s_i\in S_i,$$
where $s^\star_{-i}=\{s^\star_1,\cdots,s^\star_{i-1},s^\star_{i+1},\cdots,s^\star_n\}$.
}
\end{defi}

To enhance the players' choices, we formally define a randomized strategy so that anyone can pick a probability distribution over his set of possible strategies, which is called mixed strategy. Players evaluate the random payoff by the expected fuzzy payoff of the mixed strategy. Thus, a fuzzy Nash equilibrium of mixed strategies can be similarly defined by Definition \ref{C6}.

\subsection{Fixed point method}

Following from the total order relation of fuzzy numbers presented by Zhang et al. \cite{Zhang2019}, the continuous fuzzy payoff function and the fixed point theorem, we put forward the following theorem which refers to the existence of the fuzzy Nash equilibrium of $G_{\mathcal{F}}$ directly.

\begin{theo}\label{An}
For any fuzzy non-cooperative game $G_{\mathcal{F}}=( N,S_i,\tilde{u}_i)$, if $S_i$ is a non-empty, compact as well as convex set, then there exists a fuzzy Nash equilibrium of mixed strategies.
\end{theo}

We introduce a generalization of a fuzzy non-cooperative game named fuzzy abstract economy and define a fuzzy equilibrium of a fuzzy abstract economy. One lemma generalizing Theorem \ref{An}, gives conditions for the existence of a fuzzy equilibrium of a fuzzy abstract economy.

Let $\mathcal{H}_k \subseteq R^l$, $k=1,\cdots,n$  and $\mathcal{H}=\mathcal{H}_1\times\mathcal{H}_2\times\cdots\times\mathcal{H}_n$, i.e., $\mathcal{H}$ is the set of ordered $n$-tuple $\mathbf{a}=(\mathbf{a}_1,\mathbf{a}_2,\cdots,\mathbf{a}_n)$, where $\mathbf{a}_k\in\mathcal{H}_k$ for $k=1,\cdots,n$. For each $k$, assume that there is a fuzzy function $\tilde{f}_k$ defined over $\mathcal{H}$. Let $\mathcal{H}_{-k}= \mathcal{H}_1\times\mathcal{H}_2\times\cdots\times\mathcal{H}_{k-1}\times\mathcal{H}_{k+1}\times\cdots\times\mathcal{H}_n$, i.e., the set of ordered $(n-1)$-tuples $\mathbf{a}_{-k}=(\mathbf{a}_1,\cdots,\mathbf{a}_{k-1},\mathbf{a}_{k+1},\cdots,\mathbf{a}_n)$, where $\mathbf{a}_{k'}\in\mathcal{H}_{k'}$ for each $k'\neq k$. Assume $\mathbf{a}_{k}\in A_k(\mathbf{a}_{-k})\subseteq\mathcal{H}_{k}$ for each point $\mathbf{a}_{-k}\in \mathcal{H}_{-k}$, i.e., $A_k(\mathbf{a}_{-k})$ is a set-valued function defined for each point $\mathbf{a}_{-k}\in \mathcal{H}_{-k}$. Then the sequence
$[\mathcal{H}_1,\cdots,\mathcal{H}_n,\tilde{f}_1,\cdots,\tilde{f}_n$,
$A_1(\mathbf{a}_{-1}),\cdots,A_n(\mathbf{a}_{-n})]$ is termed a \emph{fuzzy abstract economy}.

\begin{defi}\label{C7}
\emph{
For fuzzy abstract economy
\begin{small}
$[\mathcal{H}_1,\cdots$,
$\mathcal{H}_n,\tilde{f}_1,\cdots,\tilde{f}_n,A_1(\mathbf{a}_{-1}),\cdots,A_n(\mathbf{a}_{-n})]$,
\end{small}
$\bar{\mathbf{a}}$ is a \emph{fuzzy equilibrium} point of a fuzzy abstract economy and if for all $k=1,\cdots,n$, $\bar{\mathbf{a}}_k\in A_k(\bar{\mathbf{a}}_{-k})$, it holds that
 $$\tilde{f}_k(\bar{\mathbf{a}}_{-k},\bar{\mathbf{a}}_k)\approx\max_{\mathbf{a}_k\in A_k(\bar{\mathbf{a}}_{-k})} \tilde{f}_k(\bar{\mathbf{a}}_{-k},\mathbf{a}_k).$$
}
\end{defi}

We recall some definitions in Debreu \cite{Debreu1952}. The \emph{graph} of $A_k(\mathbf{a}_{-k})$ is the set $\{\mathbf{a}\mid \mathbf{a}_k \in A_k(\mathbf{a}_{-k})\}$. The set-valued function $A_k(\mathbf{a}_{-k})$ is said to be \emph{continuous} at $ \mathbf{a}^0_{-k} $ if for any $\mathbf{a}^0_k\in A_k(\mathbf{a}^0_{-k})$ and any sequence $\{\mathbf{a}^{(n)}_{-k}\}$ converging to $\mathbf{a}^0_{-k}$, there exists a sequence $\{ \mathbf{a}^{(n)}_k\}$ converging to $\mathbf{a}^0_k$ such that for all $n$,  $\mathbf{a}^{(n)}_k\in A_k(\mathbf{a}^{(n)}_{-k})$.

Based on the total order relation of fuzzy numbers and the expected function of a fuzzy mapping, the following lemma is acquired.
\begin{lemm}\label{A3}
A fuzzy abstract economy $[\mathcal{H}_1,\cdots,\mathcal{H}_n$,
$\tilde{f}_1,\cdots,\tilde{f}_n,A_1(\mathbf{a}_{-1}),\cdots$, $A_n(\mathbf{a}_{-n})]$ has a fuzzy equilibrium point, if
\begin{enumerate}[(i)]
\item for each $k,\mathcal{H}_k$ is compact and convex, $\tilde{f}_k(\mathbf{a}_{-k},\mathbf{a}_k)$ is fuzzy continuous on $\mathcal{H}$ and fuzzy quasi-concave in $\mathbf{a}_k$;
\item for every $\mathbf{a}_{-k}$, $A_k(\mathbf{a}_{-k})$ is a continuous function whose graph is a closed set; and
\item for every $\mathbf{a}_{-k}$, the set of $A_k(\mathbf{a}_{-k})$ is convex and non-empty.
\end{enumerate}
\end{lemm}

What follows are certain assumptions concerning the consumption units in a PXE-FP. Afterwards, we can get the existence theorem of a fuzzy competitive equilibrium for $\tilde{\mathcal{E}}$.

For any good $h=1,\cdots,l$, the rate of consumption is necessarily non-negative of the agent $i=1,\cdots,m$, i.e., $x_{ih}\geq 0$.

$\mathbf{Assumption}$ I The set of consumption vectors $X_i$ available to an individual $i=1,2,\cdots,m$ is a closed convex subset of $R_+^l$, i.e., $\mathbf{x}_i \geqq \mathbf{0}$, for all $\mathbf{x}_i\in X_i$.\label{A8}

$\mathbf{Assumption}$ II For all $\mathbf{x}'_i\in X_i$, the sets $\{\mathbf{x}_i\in X_i\mid \mathbf{x}_i \precsim^i_{\mathcal{G}} \mathbf{x}'_i\}$ and $\{\mathbf{x}_i\in X_i\mid \mathbf{x}'_i\precsim^i_{\mathcal{G}} \mathbf{x}_i\}$ are closed for all $\mathbf{x}'_i\in X_i$.

This assumption ensures the continuity of $\tilde{u}_i(\mathbf{x}_i)$ demonstrated by Theorem \ref{C4}.

$\mathbf{Assumption}$ III For any $\mathbf{x}_i\in X_i$, there is $\mathbf{x}_i'\in X_i$ such that $\tilde{u}_i(\mathbf{x}_i')\succ \tilde{u}_i(\mathbf{x}_i)$.\label{A10}

This means there is no consumption vector that an individual would fuzzily prefer to all others.

$\mathbf{Assumption}$ IV If $\tilde{u}_i(\mathbf{x}_i)\succ \tilde{u}_i(\mathbf{x}_i')$ and $0<\lambda<1$, then $\tilde{u}_i[\lambda\mathbf{x}_i+(1-\lambda)\mathbf{x}_i']\succ \tilde{u}_i(\mathbf{x}_i')$.\label{B1}

This assumption corresponds to the usual assumption that the fuzzy indifference surfaces are convex in the sense that the set $\{\mathbf{x}_i\mid \mathbf{x}_i\in X_i \  \hbox{and}\  u^i_E(\mathbf{x}_i) \geq a\}$ is a convex set for any fixed real number $a$, where $u^i_E(\mathbf{x}_i)$ is the expected utility function of the fuzzy utility function $\tilde{u}_i(\mathbf{x}_i)$.

We also suppose that agent $i$ possesses an initial endowment vector $\mathbf{w}_i$ of different goods available.

$\mathbf{Assumption}$ V For some $\mathbf{x}_i\in X_i$, there exists a $\mathbf{w}_i\in R_+^l$ such that $\mathbf{x}_i<\mathbf{w}_i$. \label{B2}

This shows that any agent could exhaust his initial endowments in some feasible way and still have a positive amount of each good available for exchange in the PXE-FP.

\begin{theo}
For a PXE-FP $\tilde{\mathcal{E}}$, if $\tilde{\mathcal{E}}$ satisfies Assumptions I-V, then there is a fuzzy competitive equilibrium of $\tilde{\mathcal{E}}$.
\end{theo}
\noindent \textbf{Proof.} (i) For any agent $i$, it is supposed that ${\mathbf{x}}_{-i}$ denotes a point in $X_1\times\cdots\times X_{i-1}\times X_{i+1}\times \cdots \times X_m\times P$, i.e., ${\mathbf{x}}_{-i}$ has components as $\mathbf{x}_{i'}(i'\neq i)$, $p$. Here define
$$A_i({\mathbf{x}}_{-i})=\{\mathbf{x}_i\mid \mathbf{x}_i\in X_i, \langle\mathbf{p},\mathbf{x}_i\rangle\leq \langle \mathbf{p},\mathbf{w}_i\rangle \}.$$
Then, we consider the fuzzy abstract economy $\tilde{E}=[X_1,\cdots,X_m,P,\tilde{u}_1(\mathbf{x}_1),\cdots,\tilde{u}_m(\mathbf{x}_m),\langle \mathbf{p},\mathbf{z}\rangle, A_1({\mathbf{x}}_{-1})$,
$\cdots,A_m({\mathbf{x}}_{-m}),P]$. That is, any of $m$ consumption units chooses a vector $\mathbf{x}_i$ from $X_i$, subject to $\mathbf{x}_i\in A_i({\mathbf{x}}_{-i})$, and receives $\tilde{u}_i(\mathbf{x}_i)$; the $(m+1)$-th unit, i.e., market participant, chooses $\mathbf{p}$ from $P$ and obtains $\langle \mathbf{p},\mathbf{z}\rangle$.

Before establishing the existence of a fuzzy equilibrium point for the fuzzy abstract economy $\tilde{E}$, we intend to prove that such a fuzzy equilibrium point is also a fuzzy competitive equilibrium of PXE-FP according to  $\tilde{\mathcal{E}}$ as Definition \ref{Ay}. Obviously, Condition (1) and (2) follow immediately from the definition of a fuzzy equilibrium point of $\tilde{E}$.

Evidently, each agent spends his entire income because of the absence of saturation. To be more precise, Assumption III shows that there exists at least one consumption vector $\mathbf{x}_i'\in X_i$ such that $$\tilde{u}_i(\mathbf{x}_i')\succ\tilde{u}_i(\bar{\mathbf{x}}_i),$$ where $\bar{\mathbf{x}}_i$ is the equilibrium value of $\mathbf{x}_i$. Let $\lambda$ be an arbitrarily small positive number. On account of Assumption IV, $$\tilde{u}_i[\lambda\mathbf{x}_i'+(1-\lambda)\bar{\mathbf{x}}_i]\succ \tilde{u}_i(\bar{\mathbf{x}}_i).$$
In other words, in every neighbourhood of $\bar{\mathbf{x}}_i$, there is at least one point of $X_i$ fuzzily preferred to $\bar{\mathbf{x}}_i$. Due to Condition (1), $\langle\bar{\mathbf{p}},\bar{\mathbf{x}}_i\rangle\leq \langle\bar{\mathbf{p}},\mathbf{w}_i\rangle$. Assume
the strict inequality holds. That is, we can choose a point of $X_i$ for which the inequality still holds and which is fuzzily preferred to $\bar{\mathbf{x}}_i$, a contradiction of Condition (1). Hence, $\langle\bar{\mathbf{p}},\bar{\mathbf{x}}_i\rangle= \langle\bar{\mathbf{p}},\mathbf{w}_i\rangle$. In order to attain his equilibrium consumption plan $\bar{\mathbf{x}}_i$, agent $i$ must actually receive the total income given by the initial endowments. Thus, he cannot withhold any initial holdings of any good $h$ from the market if $p_h>0$. Since $\bar{\mathbf{z}}=(\bar{z}_1,\cdots,\bar{z}_l)$, where $\bar{z}_h=\sum\limits_{i=1}^m(x_{ih}-w_{ih})$, it is clear that
\begin{align}\label{A1}
& \langle \bar{\mathbf{p}},\bar{\mathbf{z}}\rangle=0.
\end{align}
We assume that $\mathbf{e}$ is the vector in which every component is $0$, except the $h$-th, which is $1$. Obviously $\mathbf{e}\in P$. Also, by Definition \ref{C7}, $0=\langle\bar{\mathbf{p}},\bar{\mathbf{z}}\rangle\geq \langle \mathbf{e},\bar{\mathbf{z}}\rangle=\bar{z}_h$, which means
\begin{align}\label{A2}
& \bar{\mathbf{z}}\leqq \mathbf{0}.
\end{align}
(\ref{A1}) and (\ref{A2}) together verify Condition (3). It has been shown that any fuzzy equilibrium point of $\tilde{E}$ satisfies Conditions (1)-(3), so it is a fuzzy competitive equilibrium of $\tilde{\mathcal{E}}$. The converse is obviously also true.

(ii) Unfortunately, Lemma \ref{A3} is not directly applicable to $\tilde{E}$, because the action space is not compact.
Let
\begin{equation*}
\begin{split}
&X'_{i}=\{\mathbf{x}_i\mid \mathbf{x}_i\in X_i,\ \hbox{there\ exists}\ \mathbf{x}_{i'}\in X_{i'}\ \\
  &\quad\quad\quad\quad\text{for\ each}\ i'\neq i\ \text{such that}\ \mathbf{z}\leqq \mathbf{0}\}.
\end{split}
\end{equation*}
$X'_{i}$ is the set of consumption vectors available to agent $i$ if he completely controls the economy but has to take the resource limitations into account. We plan to prove that $X'_{i}$ is bounded. Clearly, a fuzzy equilibrium point $\bar{\mathbf{x}}_i$ of $\tilde{E}$ belongs to $X'_i$.

Let $\mathbf{x}_i\in X'_{i}$. Also by Assumption I, it yields that
\begin{align*}
& \mathbf{0}\leqq \mathbf{x}_i\leqq \sum_{i=1}^m\mathbf{w}_i-\sum_{i'\neq i}\mathbf{x}_{i'},\ \mathbf{x}_{i'}\in X_{i'}.
\end{align*}
Additionally, $\mathbf{x}_{i'}\geqq \mathbf{0}$. It is true that
\begin{align*}
& \mathbf{0}\leqq \mathbf{x}_i\leqq \sum_{i=1}^m\mathbf{w}_i,
\end{align*}
which means $X'_{i}$ is bounded for all $i$.

(iii) For any $X'_i$, we can select a positive real number $c$ so that the cube $C=\{\mathbf{x}\mid |x_h|\leq c\ \hbox{for\ all}\ h\}$ contains in the interior of any $X'_{i}$. Let $X''_i=X_i\cap C$. We propose a new abstract economy $\tilde{E}'$, identical to $\tilde{E}$, except that $X_i$ is replaced by $X''_i$ everywhere. Let $A'_i(\mathbf{x}_{-i})$ be the resultant modification of $A_i(\mathbf{x}_{-i})$. It is verified that all the conditions of Lemma \ref{A3} are satisfied for $\tilde{E}'$.

From Assumption I, $X_i$ is a closed convex set and $C$ is a compact convex set. Therefore, $X''_i$ is a compact convex set. $P$ is evidently compact and convex.

For a consumption unit, the continuity and quasi-concavity of $\tilde{u}_i(\mathbf{x}_i)$ are assured by Assumption II and IV.
For the market participant, the continuity is trivial and the quasi-concavity holds for any linear function.

In addition, for the market participant, $P$ is constant and therefore trivially continuous. The closure of the graph is simply the closure of $\mathcal{H}''=X''_1\times\cdots\times X''_m\times P$. Set $P$ is certainly convex and non-empty. Moreover, for a consumption unit, the set $A'_i({\mathbf{x}}_{-i})$ is defined by a linear inequality in $\mathbf{x}_i$ and hence is certainly convex. For any $i$, let $\mathbf{x}'_i$ satisfy Assumption V, i.e., $\mathbf{x}'_i\leqq \mathbf{w}_i, \mathbf{x}'_i\in X_i$.
Since $\sum\limits_{i=1}^m (\mathbf{x}'_i-\mathbf{w}_i)\leqq\mathbf{ 0}, \mathbf{x}'_i\in X'_i$ for each $i$, it yields that $\mathbf{x}'_i\in C$. It is shown that $\mathbf{x}'_i\in A_i({\mathbf{x}}_{-i})$ for all ${\mathbf{x}}_{-i}$. On account that $A'_i({\mathbf{x}}_{-i})=A_i({\mathbf{x}}_{-i})\cap C$,
$A'_i({\mathbf{x}}_{-i})$ contains $\mathbf{x}'_i$ and therefore is non-empty.
Since the budget restraint is a weak inequality between two continuous functions of $\mathbf{p}$, it is easily seen that the graph of $A'_i({\mathbf{x}}_{-i})$ is closed.
Furthermore, from the Remark in Section 3.3.5 in Arrow and Debreu \cite{Arrow1954}, if Assumption V holds,
then $A'_i({\mathbf{x}}_{-i})$ is continuous at the point ${\mathbf{x}}_{-i}=(\mathbf{x}_1,\cdots,\mathbf{x}_{i-1},\mathbf{x}_{i+1},\cdots,\mathbf{x}_m,\mathbf{p})$.

Consequently, the existence of a fuzzy equilibrium point $(\bar{\mathbf{x}}_1,\cdots,\bar{\mathbf{x}}_m,\bar{\mathbf{p}})$ for the fuzzy abstract economy $\tilde{E}'$ has been demonstrated.

(iv) It needs to be shown that the fuzzy equilibrium point $(\bar{\mathbf{x}}_1,\cdots,\bar{\mathbf{x}}_m,\bar{\mathbf{p}})$ for the fuzzy abstract economy $\tilde{E}'$ is also a fuzzy equilibrium point for the fuzzy abstract economy $\tilde{E}$. The converse is obvious, so a fuzzy competitive equilibrium is equivalent to a fuzzy equilibrium point of $\tilde{E}'$.

From the definition of $A'_i(\mathbf{x}_{-i})$, it is true that $\langle \bar{\mathbf{p}},\bar{\mathbf{x}}_i\rangle\leq \langle \bar{\mathbf{p}},\mathbf{w}_i\rangle$. If we sum over $i$, then $\langle \bar{\mathbf{p}},\bar{\mathbf{x}}\rangle\leq \langle \bar{\mathbf{p}},\mathbf{w}\rangle$
or $\langle \bar{\mathbf{p}},\bar{\mathbf{z}}\rangle \leq 0$. For a fixed $\bar{\mathbf{z}}$, $\bar{\mathbf{p}}$ is the optimal value of the maximization problem $\langle \mathbf{p},\bar{\mathbf{z}}\rangle$ for $\mathbf{p}\in P$. By an argument similar to that used in the third paragraph of (i), we find that formula (\ref{A2}) works. From (\ref{A2}) and the definition of $X'_i$ and $C$, $\bar{\mathbf{x}}_i\in X'_i$ and $\bar{\mathbf{x}}_i$ is an interior point of $C$ for all $i$.
It is assumed that for some $\mathbf{x}'_i\in A_i(\bar{\mathbf{x}}_{-i})$, $\tilde{u}_i(\mathbf{x}'_i)\succ \tilde{u}_i(\bar{\mathbf{x}}_i)$. By Assumption IV, $$\tilde{u}_i[\lambda\mathbf{x}'_i+(1-\lambda)\bar{\mathbf{x}}_i]\succ \tilde{u}_i(\bar{\mathbf{x}}_i)\ \hbox{if}\ 0<\lambda<1.$$
However, for a sufficiently small $\lambda$,
$\lambda\mathbf{x}'_i+(1-\lambda)\bar{\mathbf{x}}_i\in C$. By the convexity of $A_i(\bar{\mathbf{x}}_{-i})$, it holds that $\lambda\mathbf{x}'_i+(1-\lambda)\bar{\mathbf{x}}_i\in A_i(\bar{\mathbf{x}}_{-i})$. Consequently, $\lambda\mathbf{x}'_i+(1-\lambda)\bar{\mathbf{x}}_i\in A'_i(\bar{\mathbf{x}}_{-i})$, which contradicts with
the definition of $\bar{\mathbf{x}}_i$ as an equilibrium value for $\tilde{E}'$. That means
$$ \bar{\mathbf{x}}_i\ \hbox{is the optimum solution of}\ \max_{\mathbf{x}_i\in A_i(\bar{\mathbf{x}}_{-i})}\tilde{u}_i(\mathbf{x}_i).$$

Meanwhile, that $\bar{\mathbf{p}}$ maximizes $\langle \mathbf{p},\bar{\mathbf{z}}\rangle$ for $\mathbf{p}\in P$ is directly implied by the definition of a fuzzy equilibrium point for $\tilde{E}'$, since the domain of $\mathbf{p}$ is the same in both fuzzy abstract economies $\tilde{E}$ and $\tilde{E}'$.

Therefore, the point $(\bar{\mathbf{x}}_1,\cdots,\bar{\mathbf{x}}_m,\bar{\mathbf{p}})$ is also a fuzzy equilibrium point for $\tilde{E}$. Additionally, as shown in the third paragraph of (i), it is a fuzzy competitive equilibrium of $\tilde{\mathcal{E}}$. \qed

\subsection{Variational approach}

Note that, the goal of any agent becomes to find his optimal consumption vector which maximizes his fuzzy utility by accomplishing the exchange of the goods in his budget set. This leads to the following optimization problem, for all $i=1,\cdots,m$ and $\mathbf{p}\in P$:
\begin{align}\label{C12}
& \tilde{u}_i(\bar{\mathbf{x}}_i)\approx\max_{\mathbf{x}_i\in B_i(\mathbf{p})}\tilde{u}_i(\mathbf{x}_i),
\end{align}
where $B_i(\mathbf{p})=\{\mathbf{x}_i\mid \mathbf{x}_i\in X_i, \langle \mathbf{p},\mathbf{x}_i\rangle\leq \langle \mathbf{p},\mathbf{w}_i\rangle\}$ is $i$'s budget set. We restrict $\mathbf{p}\in P=\{\mathbf{p}\mid \mathbf{p}\in R^l,\mathbf{p}\geqq \mathbf{0},\sum_{h=1}^l p_h=1\}$, since for all $\lambda>0$, $B_i(\mathbf{p})=B_i(\lambda\mathbf{p})$.

From the total order relation of fuzzy numbers and the expected utility function $ u^i_E(\mathbf{x}_i)$ of the fuzzy utility function $ \tilde{u}_i(\mathbf{x}_i)$, it yields that (\ref{C12}) is equivalent to saying that
\begin{align}\label{C14}
& u^i_E(\bar{\mathbf{x}}_i)=\max_{\mathbf{x}_i\in B_i(\mathbf{p})} u^i_E(\mathbf{x}_i).
\end{align}
We assume for $i=1,\cdots,m$:
 \begin{enumerate}[(i)]
 \item $u^i_{E}$ is continuous and strictly concave on $X_i$;\label{11}
 \item For any $\mathbf{x}_i\in B_i(\mathbf{p}): \nabla u^i_E(\mathbf{x}_i)\neq0$, $\forall \mathbf{p}\in P$ and $\forall \mathbf{x}_i\in \partial B_i(\mathbf{p}): \frac{\partial u^i_E(\mathbf{x}_i)}{\partial x_{is}}> 0$, when $x_{is}=0$, $\forall \mathbf{p}\in P$;\label{B7}
\item $\lim\limits_{\substack{\|\mathbf{x}_i\|\rightarrow +\infty\\\mathbf{x}_i\in B_i(\mathbf{p})}} u^i_E(\mathbf{x}_i)=-\infty$; and \label{B8}
\item Any agent is endowed with a positive quantity of at least one good, i.e., there exists a good $h$ such that $w_{ih}>0$ for all $i=1,\cdots,m$.\label{22}
\end{enumerate}
Under Assumptions (\ref{11})-(\ref{22}), for all $i=1,\cdots,m$, the maximization problem (\ref{C14}), i.e., (\ref{C12}), has a unique solution $\bar{\mathbf{x}}_i(\mathbf{p})$ for each $\mathbf{p}\in P$, denoted by $\bar{\mathbf{x}}_i$.

Therefore, the fuzzy competitive equilibrium of Definition \ref{Ay} is equivalent to the following statement:

\begin{defi}
\emph{
For PXE-FP $\tilde{\mathcal{E}}$, let $\bar{\mathbf{p}}\in P$ and $\bar{\mathbf{x}}\in B(\bar{\mathbf{p}})=\prod\limits_{i=1}^m B_i(\bar{\mathbf{p}})$. The pair $(\bar{\mathbf{p}},\bar{\mathbf{x}})\in P\times B(\bar{\mathbf{p}})$ is a \emph{fuzzy competitive equilibrium} of $\tilde{\mathcal{E}}$ if and only if
\begin{equation}\label{B6}
 \tilde{u}_i(\bar{\mathbf{x}}_i)\approx\max_{\mathbf{x}_i\in B_i(\bar{\mathbf{p}})}\tilde{u}_i(\mathbf{x}_i),\ \hbox{for all}\ i=1,\cdots,m,
\end{equation}
and
\begin{align*}
& z_h=\sum\limits_{i=1}^m (\bar{x}_{ih}-w_{ih})\leq 0, \ \hbox{for any}\ h=1,\cdots,l.
\end{align*}
}
\end{defi}
Analogous to (\ref{C12}), (\ref{B6}) is equivalent to saying that
\begin{align*}
& u^i_E(\bar{\mathbf{x}}_i)=\max_{\mathbf{x}_i\in B_i(\bar{\mathbf{p}})} u^i_E(\mathbf{x}_i).
\end{align*}
Consequently, from Theorem $1$ in Anello et al. \cite{Anello2010}, it is obvious that the pair $(\bar{\mathbf{p}},\bar{\mathbf{x}})\in P\times B(\bar{\mathbf{p}})$ is a fuzzy competitive equilibrium of a PXE-FP if and only if it is a solution to the following quasi-variational inequality:
\begin{align}\label{AK}
& \sum_{i=1}^m\langle-\nabla u^i_E(\bar{\mathbf{x}}_i),(\mathbf{x}_i-\bar{\mathbf{x}}_i)\rangle- \nonumber \\
 &\quad\quad\quad\quad\langle\sum_{i=1}^m(\bar{\mathbf{x}}_i-\mathbf{w}_i),(\mathbf{p}-\bar{\mathbf{p}})\rangle\geq0,
\end{align}
for any $(\mathbf{p},\mathbf{x})\in P \times B(\bar{\mathbf{p}})$.

Donato et al. \cite{Donato2008} gave the existence theorem of solutions to quasi-variational inequality problem (\ref{AK}) in Theorem $4$.

$(\bar{\mathbf{p}},\bar{\mathbf{x}})\in P \times B(\bar{\mathbf{p}})$ is a solution of (\ref{AK}) if and only if
for all $i=1,\cdots,m$, $\bar{\mathbf{x}}_i(\mathbf{p})$ is a solution to
\begin{equation}\label{AM}
\langle-\nabla  u^i_E(\bar{\mathbf{x}}_i),(\mathbf{x}_i-\bar{\mathbf{x}}_i)\rangle\geq 0,\ \hbox{for all}\ \mathbf{x}_i\in B_i(\mathbf{p}).
\end{equation}
and $\bar{\mathbf{p}}$ is the solution to
\begin{equation}\label{AL}
\langle-\sum_{i=1}^m(\bar{\mathbf{x}}_i-\mathbf{w}_i),(\mathbf{p}-\bar{\mathbf{p}})\rangle\geq0,\ \hbox{for all}\ \mathbf{p}\in P.
\end{equation}
Observe that, when the operator $-\nabla  u^i_E(\bar{\mathbf{x}}_i)$ is strongly monotone, variational inequality (\ref{AM}) has a unique solution. If $\bar{\mathbf{x}}_i$ is a continuous function, then variational inequality problem (\ref{AL}) admits a solution $\bar{\mathbf{p}}\in P$, seeing that $P$ is closed, convex and bounded.

Accordingly, we get the following theorem about the existence of fuzzy equilibrium solutions immediately by an associated quasi-variational inequality.
\begin{theo}
The pair $(\bar{\mathbf{p}},\bar{\mathbf{x}})\in P\times B(\bar{\mathbf{p}})$ is a fuzzy competitive equilibrium of a PXE-FP if and only if $(\bar{\mathbf{p}},\bar{\mathbf{x}})$ is a solution to the quasi-variational inequality (\ref{AK}).
\end{theo}
The following example will illustrate how to research the fuzzy competitive equilibrium by a related quasi-variational inequality.
\begin{exam}
\emph{
We consider a market consisting of two different goods, denoted by good $h=1,2$, and two agents, i.e., agent $i=1,2$. Each agent is endowed with an initial vector $\mathbf{w}_i=(w_{i1},w_{i2})$. The consumption vector of any agent is $\mathbf{x}_i=(x_{i1},x_{i2})$. It is assumed that each commodity is sold and purchased at only one price and the price vector is $\mathbf{p}=(p_1,p_2)$ satisfying $p_1+p_2=1$. Now we suppose that each agent has a fuzzy utility function defined as follows:
\begin{align*}
&\tilde{u}_i(x_{i1},x_{i2})=\tilde{-}\lfloor0,\frac{1}{2},\frac{1}{2},1\rfloor x^2_{i1}\tilde{-}\lfloor0,\frac{1}{3},\frac{2}{3},1\rfloor x^2_{i2}\\
 &\tilde{-} \lfloor2b_{i1},b_{i1},b_{i1},0\rfloor x_{i1}\tilde{-}\lfloor2b_{i2},\frac{3}{2}b_{i2},\frac{1}{2}b_{i2},0\rfloor x_{i2}\\
&\tilde{\pm}\lfloor2c_{i},c_{i},c_{i},0\rfloor.
\end{align*}
It is easy to work out that
$$u^i_E(x_{i1},x_{i2})=-\frac{1}{2}x^2_{i1}-\frac{1}{2}x^2_{i2}-b_{i1}x_{i1}-b_{i2}x_{i2}\pm c_i.$$
For agent $i=1,2$, $-\nabla u^i_E(\mathbf{x}_i)=(x_{i1}+b_{i1},x_{i2}+b_{i2})$, we fix $\mathbf{p}\in P$ and find $\bar{\mathbf{x}}_i\in B_i(\mathbf{p})=\{\mathbf{x}_i=(x_{i1},x_{i2})\mid p_1(x_{i1}-w_{i1})+p_2(x_{i2}-w_{i2})\leq 0\}$ such that for all $\mathbf{x}_i\in B_i$,
\begin{equation}\label{AN}
(\bar{x}_{i1}+b_{i1})(x_{i1}-\bar{x}_{i1})+(\bar{x}_{i2}+b_{i2})(x_{i2}-\bar{x}_{i2})\geq0.
\end{equation}
Notice that $x_{ij}$ is the function of $\mathbf{p}$. Since the operator $-\nabla u^i_E(\mathbf{x}_i)$ is strongly monotone, there exists a unique solution $\bar{\mathbf{x}}_i$ to the variational inequality. Assumption (\ref{B7}) is verified if we assume $-b_{ih}>w_{ih}$. Obviously, the solution to (\ref{AN}) lies in the following set:
\begin{equation}\label{AO}
\{\mathbf{x}_i\in R^2_+:p_1(x_{i1}-w_{i1})+p_2(x_{i2}-w_{i2})=0\}.
\end{equation}
}
\emph{
Moreover, search for the solution $\bar{\mathbf{p}}=(\bar{p}_1,\bar{p}_2)\in P$ such that for all $\mathbf{p}=(p_1,p_2)\in P$,
\begin{align}\label{AX}
& -z_1(p_1-\bar{p}_1)+(-z_2)(p_2-\bar{p}_2)\geq0,
\end{align}
where $z_h=(\bar{x}_{1h}-w_{1h})+(\bar{x}_{2h}-w_{2h})$ is the aggregate excess demand function of any good $h=1,2$.
Since $p_1+p_2=1$, from (\ref{AX}), it yields that
\begin{align}\label{AY}
& (z_1-z_2)(p_2-\bar{p}_2)\geq0, \ \hbox{for} \ p_2\in [0,1].
\end{align}
The solution of (\ref{AY}) is identical to the solution of the following system:
\begin{equation} \label{AZ}
 \left\{%
 \begin{array}{ll}
 z_1-z_2=0, \\
\bar{p}_1=1-\bar{p}_2, \bar{p}_2\geq0.
\end{array}%
\right.
\end{equation}
We discuss the solution to (\ref{AN}) and (\ref{AY}) by cases as follows:
}

\emph{
\textbf{Case1.} Assume that $p_1,p_2\neq 0$. For agent $i=1,2$, by (\ref{AO}), it is true that
$$ x_{i2}=w_{i2}-\frac{p_1}{p_2}(x_{i1}-w_{i1}) \ \hbox{and} \ 0\leq x_{i1}\leq w_{i1}+w_{i2}\frac{p_2}{p_1}.$$
Furthermore, from (\ref{AN}), it is easily seen that
\begin{equation*}
\begin{split}
&\left[\frac{p^2_1+p^2_2}{p^2_2}\bar{x}_{i1}-(w_{i2}+b_{i2})\frac{p_1}{p_2}-w_{i1}(\frac{p_1}{p_2})^2+b_{i1}\right]\\
&\quad\quad\quad\quad(x_{i1}-\bar{x}_{i1})\geq0.
\end{split}
\end{equation*}
Accordingly, solving (\ref{AN}) is equivalent to solving the system as follows:
\begin{small}
\begin{equation} \label{AP}
 \left\{%
 \begin{array}{ll}
\frac{p^2_1+p^2_2}{p^2_2}\bar{x}_{i1}-(w_{i2}+b_{i2})\frac{p_1}{p_2}-w_{i1}(\frac{p_1}{p_2})^2+b_{i1}=0,\\
0\leq \bar{x}_{i1}\leq w_{i1}+w_{i2}\frac{p_2}{p_1},\bar{x}_{i2}=w_{i2}-\frac{p_1}{p_2}(x_{i1}-w_{i1}).
\end{array}%
\right.
\end{equation}
\end{small}
The solution to (\ref{AP}) is
\begin{small}
\begin{equation} \label{AQ}
\left\{ {\begin{array}{*{20}{c}}
\bar{x}_{i1} =& \frac{{p_2^2}}{{p_1^2 + p_2^2}}\left[ {w_{i1}{{(\frac{{{p_1}}}{{{p_2}}})}^2} +(w_{i2}+b_{i2}) \frac{{{p_1}}}{{{p_2}}} -b_{i1}} \right],\\
{}\\
\bar{x}_{i2} =& \frac{{p_1^2}}{{p_1^2 + p_2^2}}\left[ {w_{i2}{{(\frac{{{p_2}}}{{{p_1}}})}^2} +(w_{i1}+b_{i1})\frac{{{p_2}}}{{{p_1}}} -b_{i2}} \right],
\end{array}} \right.
\end{equation}
\end{small}
under the condition that for $(p_1,p_2)\in P$,
\begin{equation}\label{33}
 \left\{%
 \begin{array}{ll}
  {w_{i1}{{(\frac{{{p_1}}}{{{p_2}}})}^2} +(w_{i2}+b_{i2}) \frac{{{p_1}}}{{{p_2}}} -b_{i1}}\geq 0, \\
{w_{i2}{{(\frac{{{p_2}}}{{{p_1}}})}^2} +(w_{i1}+b_{i1})\frac{{{p_2}}}{{{p_1}}} -b_{i2}}\geq 0.
\end{array}%
\right.
\end{equation}
}

\emph{
\textbf{Subcase1.} If condition (\ref{33}) holds for any agent $i=1,2$, then the solution to the variational inequality (\ref{AN}) is (\ref{AQ}). Combining formulas (\ref{AZ}) and (\ref{AQ}), it is clearly known that
\begin{equation} \label{Aa}
 \left\{%
 \begin{array}{ll}
 \frac{Bp^2_1+(B-A)p_1p_2-Ap^2_2}{p^2_1+p^2_2}=0, \\
p_2=1-p_1.
\end{array}%
\right.
\end{equation}
Consequently, the solution to (\ref{Aa}), i.e., (\ref{AY}) is
\begin{equation*}
 \left\{%
 \begin{array}{ll}
 p_1=\frac{A}{A+B}, \\
 {}\\
p_2=\frac{B}{A+B},
\end{array}%
\right.
\end{equation*}
where $A=w_{11}+b_{11}+w_{21}+b_{21}, B=w_{12}+b_{12}+w_{22}+b_{22}$.
}

\emph{
 \textbf{Subcase2.} If system (\ref{AP}) does not have any solution, then we find the solution to (\ref{AN}) on the boundary of the set $\{\mathbf{x}_i\in R^2_+:p_1(x_{i1}-w_{i1})+p_2(x_{i2}-w_{i2})=0\}$, which is either $\bar{x}_{i1}=0$ or $\bar{x}_{i2}=0$. The pair
\begin{equation} \label{Ab}
 \left\{%
 \begin{array}{ll}
 \bar{x}_{i1}=0, \\
\bar{x}_{i2}=w_{i2}+w_{i1}\frac{p_1}{p_2},
\end{array}%
\right.
\end{equation}
is the solution to variational inequality (\ref{AN}) if and only if
\begin{align}\label{Ac}
& w_{i1}(\frac{p_1}{p_2})^2+(w_{i2}+b_{i2})\frac{p_1}{p_2}-b_{i1}<0,
\end{align}
for $(p_1,p_2)\in P$.}

\emph{If (\ref{Ac}) does not hold,  it is true that
\begin{equation} \label{Ad}
 \left\{%
 \begin{array}{ll}
 \bar{x}_{i1}=w_{i1}+w_{i2}\frac{p_2}{p_1}, \\
\bar{x}_{i2}=0,
\end{array}%
\right.
\end{equation}
is the solution to variational inequality (\ref{AN}) if and only if
\begin{align}\label{Ae}
& w_{i2}(\frac{p_2}{p_1})^2+(w_{i1}+b_{i1})\frac{p_2}{p_1}-b_{i2}<0,
\end{align}
for $(p_1,p_2)\in P$.}

\emph{Note that solution (\ref{Ab}) or (\ref{Ad}) in this case is continuous in $P$.
}

\emph{
If condition (\ref{Ac}) holds, for any agent $i=1,2$, the solution to (\ref{AP}) is (\ref{Ab}). Under this situation, solving (\ref{AY}) is equivalent to solving the system
\begin{equation} \label{Af}
 \left\{%
 \begin{array}{ll}
 1+\frac{p_1}{p_2}=0, \\
p_1=1-p_2\geq 0,p_2>0.
\end{array}%
\right.
\end{equation}
It is found that (\ref{Af}) has no solution, which shows the solution to (\ref{AY}) in the boundary of $P$, i.e., $\bar{\mathbf{p}}=(0,1)$.
}

\emph{
If (\ref{Ae}) holds for every agent $i=1,2$, the solution to (\ref{AP}) is (\ref{Ad}). Under this circumstance, the solution of (\ref{AY}) is the same as the solution to the following system
\begin{equation} \label{Ak}
 \left\{%
 \begin{array}{ll}
 1+\frac{p_2}{p_1}=0, \\
p_1=1-p_2> 0,p_2\geq0.
\end{array}%
\right.
\end{equation}
It is true that (\ref{Ak}) has no solution, which implies that the solution to (\ref{AY}) lies in the boundary of $P$, i.e., $\bar{\mathbf{p}}=(1,0)$.
}

\emph{
\textbf{Case2.} Given that $p_1=0$ and $p_2=1$, the budget set of agent $i=1,2$ is $B_i(0,1)=\{\mathbf{x}_i\in R^2_+\mid x_{i2}\leq w_{i2}\}$. Hence, the solution to (\ref{AN}) is
$(\bar{x}_{i1}(0,1),\bar{x}_{i2}(0,1))=(-b_{i1},w_{i2})$.
Moreover, $\mathbf{p}=(0,1)$ is the solution to (\ref{AY}) if and only if $z_1-z_2<0$. But from $(\bar{x}_{i1}(0,1),\bar{x}_{i2}(0,1))=(-b_{i1},w_{i2})$, we get that
$z_1-z_2=-b_{11}-w_{11}-b_{21}-w_{21}>0$, which contradicts with $z_1-z_2<0$. In other words, $\mathbf{p}=(0,1)$ is not the solution to (\ref{AY}).
}

\emph{
\textbf{Case3.} Provided that $p_1=1$ and $p_2=0$, the budget set of agent $i=1,2$ is $B_i(1,0)=\{\mathbf{x}_i\in R^2_+\mid x_{i1}\leq w_{i1}\}$. As a consequence, the solution to (\ref{AN}) is
$(\bar{x}_{i1}(1,0),\bar{x}_{i2}(1,0))=(w_{i1},-b_{i2})$.
Furthermore, $\mathbf{p}=(1,0)$ is the solution to (\ref{AY}) if and only if $z_1-z_2>0$. Nevertheless, from $(\bar{x}_{i1}(1,0),\bar{x}_{i2}(1,0))=(w_{i1},-b_{i2})$, we obtain that
$z_1-z_2=b_{12}+w_{12}+b_{22}+w_{22}<0$, which contradicts with $z_1-z_2>0$. That is, $\mathbf{p}=(1,0)$ is not the solution to (\ref{AY}).
}

\emph{
In a word, $p_1=\frac{A}{A+B},p_2=\frac{B}{A+B}$ is the unique solution to (\ref{AY}). As a result, the fuzzy competitive equilibrium of the PXE-FP is $(\bar{\mathbf{p}},\bar{\mathbf{x}})$, where $$\bar{\mathbf{p}}=(\frac{A}{A+B},\frac{B}{A+B}),$$
\begin{tiny}
\begin{equation*}
\bar{\mathbf{x}}=\begin{pmatrix}
\frac{w_{11}A^2+(w_{12}+b_{12})AB-b_{11}B^2}{A^2+B^2}& \frac{w_{12}B^2+(w_{11}+b_{11})AB-b_{12}A^2}{A^2+B^2}\\
{}\\
\frac{w_{21}A^2+(w_{22}+b_{22})AB-b_{21}B^2}{A^2+B^2} & \frac{w_{22}B^2+(w_{21}+b_{21})AB-b_{22}A^2}{A^2+B^2}
\end{pmatrix}.
\end{equation*}
\end{tiny}
}

\emph{
We explain the solution of the quasi-variational inequality from the point of economics as follows:
\begin{enumerate}[(i)]
\item The supply of good $h=1,2$ equals to the demand following from
\begin{small}
\begin{eqnarray*}
\bar{x}_{11}+\bar{x}_{21}& = &\frac{(w_{11}+w_{21})A^2+AB^2-(b_{21}+b_{11})B^2}{A^2+B^2}\\
& = &\frac{(w_{11}+w_{21})A^2+(w_{11}+w_{21})B^2}{A^2+B^2}\\
& = &w_{11}+w_{21}.
\end{eqnarray*}
\end{small}
In the same way, it holds that $\bar{x}_{12}+\bar{x}_{22}=w_{12}+w_{22}.$
\item At the equilibrium price, both agents can afford their allocation for their given initial endowments, i.e.,
\begin{tiny}
\begin{align*}
&\bar{p}_1(\bar{x}_{11}-w_{11})+\bar{p}_2(\bar{x}_{12}-w_{12})=\\
&\frac{A}{A+B}\left[\frac{w_{11}A^2+(w_{12}+b_{12})AB-b_{11}B^2}{A^2+B^2}-w_{11}\right]\\
&+\frac{B}{A+B}\left[\frac{w_{12}B^2+(w_{11}+b_{11})AB-b_{12}A^2}{A^2+B^2}-w_{12}\right]\\
&=\frac{A}{A+B}\left[\frac{(w_{12}+b_{12})AB-(b_{11}+w_{11})B^2}{A^2+B^2}\right]\\
&+\frac{B}{A+B}\left[\frac{(w_{11}+b_{11})AB-(b_{12}+w_{12})A^2}{A^2+B^2}\right]=0.
\end{align*}
\end{tiny}
Analogously, $\bar{p}_1(\bar{x}_{21}-w_{21})+\bar{p}_2(\bar{x}_{22}-w_{22})=0$.
\item Agent $i=1,2$ fuzzily weakly prefers the consumption bundle $(\bar{x}_{i1},\bar{x}_{i2})$ to the initial endowment vector $(w_{i1},w_{i2})$. That is,
\begin{tiny}
\begin{align*}
&E(\tilde{u}_1(\bar{x}_{11},\bar{x}_{12}))-E(\tilde{u}_1(w_{11},w_{12}))=\\
&\frac{w^2_{11}(A^2+B^2)^2-[w_{11}A^2+(w_{12}+b_{12})A B-b_{11}B^2]^2}{2(A^2+B^2)^2}\\
&-\frac{w^2_{12}(A^2+B^2)^2[w_{12}B^2+(w_{11}+b_{11})A B-b_{12}A^2]^2}{2(A^2+B^2)^2}\\
&-\frac{b_{11}[w_{11}A^2+(w_{12}+b_{12})A B-b_{11}B^2]}{A^2+B^2}\\
&-\frac{b_{12}[w_{12}B^2+(w_{11}+b_{11})A B-b_{12}A^2]}{A^2+B^2}\\
&+b_{11}w^2_{11}+b_{12}w^2_{12}\\
&=\frac{[(w_{11}+b_{11})B-(w_{12}+b_{12})A]^2}{2(A^2+B^2)}\geq 0.
\end{align*}
\end{tiny}
\end{enumerate}
}

\emph{
Hence, it is true that $\tilde{u}_1(\bar{x}_{11},\bar{x}_{12})\succcurlyeq\tilde{u}_1(w_{11},w_{12})$.
Moreover, if the equality holds, $(w_{11}+b_{11})B=(w_{12}+b_{12})A$. At this time, $\bar{x}_{1h}=w_{1h}$, $h=1,2$, which means the initial endowment vector for agent $1$ is optimal.
Similarly, we can get that $\tilde{u}_2(\bar{x}_{21},\bar{x}_{22}) \succcurlyeq\tilde{u}_2(w_{21},w_{22})$. It is figured out that two goods are distributed efficiently between two agents after an exchange of goods.
}
\end{exam}

\section{Conclusion}
In this paper, we established a fuzzy binary relation to evaluate the fuzzy preference or indifference of alternative consumption vectors and proved that there exists a continuous fuzzy order-preserving function on the consumption set under some assumptions. Furthermore, the existence result of fuzzy competitive equilibrium for the PXE-FP was provided through two different methods. Therefore, when any agent's attitude is vague, we can find out the redistribution of the agents and the price vector of the goods in view of the model of pure exchange economy with fuzzy preference proposed in our paper.

We only proved the existence of the competitive equilibrium for the PXE-FP under some assumptions in this paper. Future research on the PXE-FP will be done on, for example, the uniqueness and the stability of the fuzzy competitive equilibrium. The latter study would require the specification of the dynamics of a competitive market with fuzzy preferences. Finally, the existence and the uniqueness of the equilibrium could be verified by applying the generalized linear discrete-time system with fuzzy dynamic PXE-FP. A concrete dynamic PXE-FP simulation model could also be provided to confirm the results.

\section*{Acknowledgements}
We are indebted to Arantza Estévez-Fernández for the help of modifying the language and we are grateful to two anonymous referees for some helpful remarks. The research has been supported by the National Natural Science Foundation of China (Grant Nos. 71571143).

\end{document}